# The Readability of Tweets and their Geographic Correlation with Education

**James R. A. Davenport**
University of Washington
Seattle, WA
jrad@astro.washington.edu

**Robert DeLine**
Microsoft Research
Redmond, WA
rdeline@microsoft.com

**ABSTRACT**
Twitter has rapidly emerged as one of the largest worldwide venues for written communication. Thanks to the ease with which vast quantities of tweets can be mined, Twitter has also become a source for studying modern linguistic style. The readability of text has long provided a simple method to characterize the complexity of language and ease that documents may be understood by readers. In this note we use a modified version of the Flesch Reading Ease formula, applied to a corpus of 17.4 million tweets. We find tweets have characteristically more difficult readability scores compared to other short format communication, such as SMS or chat. This linguistic difference is insensitive to the presence of "hashtags" within tweets. By utilizing geographic data provided by 2% of users, joined with "ZIP Code Tabulation Area" (ZCTA) level education data from the U.S. Census, we find an intriguing correlation between the average readability and the college graduation rate within a ZCTA. This points towards a difference in either the underlying language, or a change in the type of content being tweeted in these areas.

**Author Keywords**
Design; Human Factors; Measurement; Linguistics

**ACM Classification Keywords**
H.5.m. Information interfaces and presentation (e.g., HCI)

## INTRODUCTION
The communication revolution from microblogging platforms such as Twitter.com has ushered in new ways to quantitatively study language, communication, and social interaction on previously impossible scales. We are fundamentally interested in the regional variations of language and communication, both in its complexity and in content, and how these variations are borne differently in new mediums such as Twitter.

The "readability" of text has long been used to predict the difficulty people will have in understanding written content. A variety of approaches have historically been used to calculate readability scores. These range from counting the density of previously identified "difficult" words, to various schemes of counting the number of syllables per word and words per sentence. Many public agencies require that laws and official documents conform to thresholds of readability or "Reading Ease" scores. This is intended to promote and enable comprehension of such material by the general public.

In this note we present a novel use of traditional readability measurements in studying the language of Twitter for a large sample of tweets. We first outline our definition of readability and the samples of data used. We then investigate the impact of hashtags on this metric. Finally we describe a correlation between education and the language being used. This research indicates that despite its abbreviated format, the language on Twitter may trace underlying differences in communication within society, which should be considered when tweeting information to the general public.

## RELATED WORK
Twitter has been extensively studied, both for tweet content and social network analysis. An important and laborious step in understanding tweets comes from examining their content [1]. The structure and formality of language in Twitter has been characterized by Hu *et al* [7].

Readability and linguistic style can play a role in the way scientists read and cite scientific papers [5]. Kim *et al*. [8] have demonstrated that customizing the readability of search results based on a user's reading comprehension level can enhance search engine functionality for children and people with lower levels of education. Public announcements and discourse on Twitter may also benefit from tailoring content based on a user's reading level.

## MEASURING READING EASE
For this work we used a traditional readability metric, the Flesch reading ease formula [4]. We chose this metric because it bases the readability score primarily on word and sentence length, which is are simple quantities to measure within each tweet. Other readability metrics require taking measurements over large amounts of text (e.g. the Gunning fog index) [6], or counting matches to a corpus of "difficult words" (e.g. the Dale-Chall formula) [3], which may be inaccurate due to misspellings or abbreviations.

Our version of the reading ease (hereafter RE) equation was slightly modified to accommodate Twitter's short format. The standard Flesch formula is:

$$RE = 206.835 - 1.015 \left(\frac{\#Words}{\#Sentences}\right) - 84.6 \left(\frac{\#Syllables}{\#Words}\right)$$

This produces values nominally in the range [0,122], with higher RE scores indicating greater ease in readability, and lower scores indicating difficult or complex writing. Values below this range are also possible for very complex texts.

Because tweets are inherently short and punctuation is often unconventional, we treat each tweet as having a single sentence. We followed standard conventions to define syllables within words: split words on vowels, pairs of vowels count as a single split. For words ending in –e (except those ending with –le), –es, or –ed, we subtracted a syllable. Any words that started with "http" or "@" were removed, thus excluding hyperlinks and usernames from the calculation. Tweets with a null score (e.g. those only containing a hyperlink) were discarded. We calculated our RE metric both with and without words that started with "#", colloquially known as "hashtags". We discuss their effect on the readability later on.

**OUR SAMPLE OF TWEETS**

To build our sample of tweets to measure readability from, we first obtained 49 million tweets directly from the Twitter API on 2013 July 3. From these we selected tweets marked with language code $lang = "en"$, generating a sample of 17.4 million English language tweets suitable for measuring RE on. In Figure 1 we show the distribution of RE measurements (with hashtags removed) for this sample. The mean RE was calculated to be 50.803±0.006, where the uncertainty quoted is the standard error.

**COMPARING TWITTER TO SMS AND CHAT**

Since no previous sample of RE measurements for a large sample of tweets are available to compare our analysis to, we selected two other short format written datasets to replicate our analysis on. We used the NUS[1] corpus of 51k English SMS (text messages), which traditionally have a limit of 160 characters (compared to 140 characters for Twitter), as well as the standard NPS corpus of ~10k chat room messages[2]. Due to their similarly compressed format and style, these datasets have been used in the past as comparisons to Twitter to study the formality of language [7]. In Figure 1 we also show the distribution of RE scores for these comparison datasets. The average RE for chat room messages was higher than Twitter, with a mean score of 54.0±0.3. We find the average RE for SMS was 88.2±0.1, much higher (i.e. easier to read) than that of Twitter despite the additional 20 characters available. We speculate this may be due to the difference in subject matter and content between tweets and SMS (e.g. public sharing of news versus personal updates).

---
[1] http://wing.comp.nus.edu.sg:8080/SMSCorpus/

[2] http://faculty.nps.edu/cmartell/NPSChat.htm

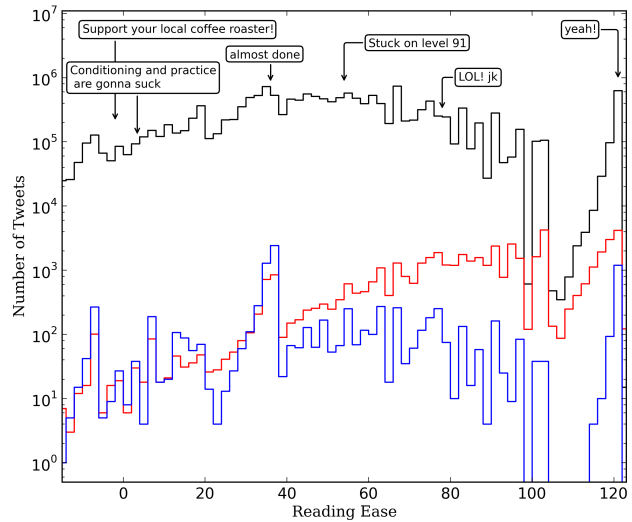

**Figure 1.** Histograms of RE for our three samples, 17.4 million tweets (black), the NUS SMS corpus (red), and the NPS chat corpus (blue). Higher RE values are easier to read. Example tweets are shown, with arrows indicating their RE score.

A few distinct features are apparent in all three distributions in Figure 1. An exponential spike is seen with a peak at RE=122, corresponding to a large number of messages with an average of 1 syllable per word (e.g. "lol" or "omg"). Similarly, another spike is seen at RE=36 due to messages with an average of 2 syllables per word (e.g. "haha"). The underlying smooth distribution of RE in Figure 1 is the result of more "normal" sentences, and drives the mean RE scores.

**THE EFFECT OF HASHTAGS ON READING EASE**

An integral part of the Twitter language is the presence of "hashtags", words that start with #. These keywords are dynamic in their style, contextual, searchable, and often found at the end of a tweet. They can be relevant to an event or place (e.g. "#Election"), or can be used to convey emotion or value-added information (e.g. "I love #Seattle"). A single tweet may contain many such tags, and hashtags may be constructed of many words (e.g. "#NeedToDoLaundry"), though use of capitalization for each word within a hashtag is not universally adopted. Correctly parsing hashtags in to individual words is therefore beyond the scope of this work.

For our analysis we calculated RE scores both including hashtags (after removing the #) and removing hashtags entirely from the tweets. While the presence of hashtags can greatly affect the RE score for a single tweet, we find the ensemble distribution of RE is virtually unchanged. From our sample of 17.4 million tweets, only 2.8 million (16%) contain one or more hashtags.

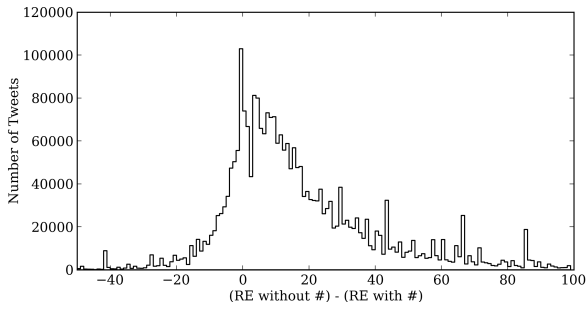

**Figure 2. Histogram of difference in RE due to including hashtags (#) in the RE calculation for the 2.8 million tweets that contain at least one hashtag.**

In Figure 2 we show the difference in RE score that resulted from including hashtags in our computation for the 2.8 million tweets that contained at least one hashtag. The median difference in RE in Figure 2 was +11.1, indicating that including hashtags in the RE calculations overall makes tweets *harder* to read (lower RE) score. However, this change in RE due to hashtags only produced a small shift in the ensemble average for our sample of 17.4 million tweets, changing the mean RE score from 50.803±0.006 to 48.426±0.006. For the rest of this work we utilize our initial RE measurement that excluded hashtags.

**CORRELATING READABILITY WITH DEMOGRAPHICS**

From our primary sample of 17.4 million tweets, only a small portion contained geographic information (latitude and longitude coordinates) with the tweet. Here we only consider geo-data from the Twitter API for tweets, rather than "location" strings provided by the users. This geo-data comes primarily from smart phones, and users must opt-in to including it. We identified 336,126 tweets (2%) with geographic data included. Previous large scale studies of Twitter geo-data have found user opt-in rates ranging from 0.2% to 1% [2]. This may indicate growth in the adoption of mobile geo-tagging.

The mean RE score for these geo-tagged tweets was 51.43±0.05, somewhat higher (easier to read) than for the Twitter sample as a whole. This may be due to the increased burden for text input on mobile devices, which causes users to further abbreviate their linguistic style. However, it is notable that this average RE score is still lower (more complex language) than that of the SMS and chat corpuses.

We utilized the geo-data from these tweets to search for correlations between RE scores and geographically indexed demographic data. The U.S. Census Bureau provides a wealth of demographic data broken down by various geographic regions. For our study we used Census data indexed by "ZIP Code Tabulation Area" (hereafter ZCTA),

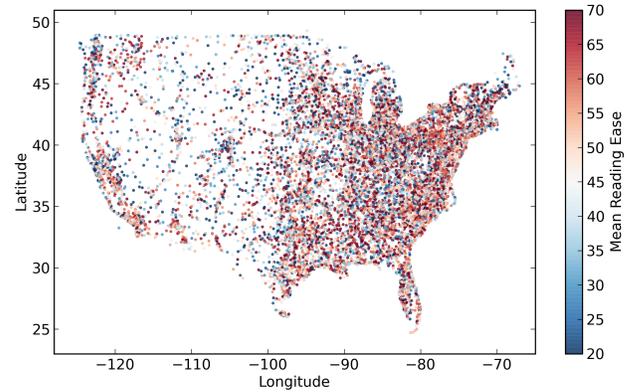

**Figure 3. Map of average RE scores (excluding hashtags) for the 5151 individual ZCTAs with at least 10 tweets.**

which include 33,121 regions that roughly correspond to U.S. postal ZIP codes[3].

Of the 336k tweets with geo-data in our sample, 227,526 fell within 10 degrees of a ZCTA center, nominally placing these tweets within the USA, and excluding other English speaker countries such as Canada. We grouped theses tweets by their corresponding ZCTA, finding 5151 ZCTA's containing 10 or more tweets. For each ZCTA we determined the mean RE score (excluding hashtags). In Figure 3 we present the spatial distribution of these ZCTAs for the continental United States, color coded by their mean RE scores. No significant large scale geographic trend with RE was found using the relatively fine scaled ZCTAs. We did not study average RE scores over larger geographic regions such as metropolitan areas or States.

The U.S. Census Bureau's 2007-2011 American Community Survey[4] (ACS) produced population and demographic information, as well as education level attainment data. These data are available broken down by ZCTA. We used table S1501 from the ACS to study the correlation between regional educational-attainment and mean RE scores.

We first investigated the percentage of the ZCTA's population with a high school (or equivalent) degree as a possible indicator for average RE score. However, no significant correlation was found between the two due to the small dynamic range present in high school graduation rates. For the 5151 ZCTA's present in our sample, 75% had high school graduation rates of 80% or higher.

---

[3] http://www.census.gov/geo/reference/zctas.html

[4] http://www.census.gov/acs/www/

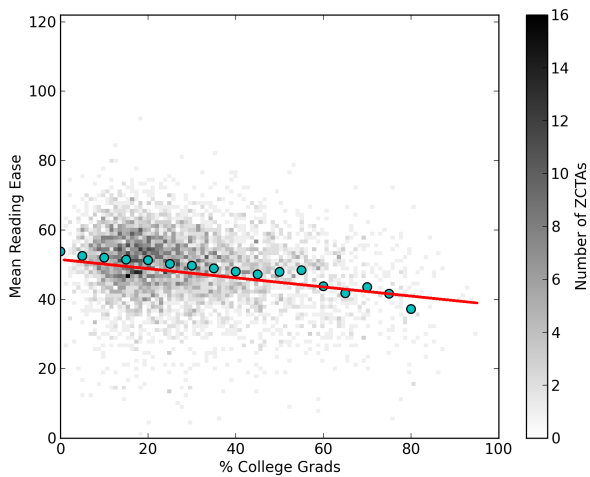

**Figure 4. Density plot of average reading ease versus college graduation rate for 5151 individual ZCTAs with at least 10 tweets. An error-weighted least squares fit to this data is provided (red line), along with the median values of average reading ease in bins (blue circles).**

We then studied the fraction of a ZCTA's population holding a bachelors (4 year) college degree. In Figure 4 we show a density map of the average RE score versus percentage of population with college degree. A significant correlation between the two is present. For each ZCTA we estimated the uncertainty on the mean RE using the standard error. We fit the distribution in Figure 4 with an error-weighted least-squares linear regression, which provided a slope of $\beta = -0.132 \pm 0.009$. For independent comparison we also computed the median of the average RE scores in bins as a function of college graduation %, shown as blue circles.

**THREATS TO VALIDITY**

We made no correction or accomodation for abbreviations. As such we consider this an empirical readability score. In many cases our score may underpredict the linguistic complexity of the text, especially for those tweets employing many abbreiations to compress a large amount of information in to 140 charaters. Conversely our metric may grossly overpredict complexity (lower RE score) in some cases, for example due to long and sometimes complicated non-words (e.g. "LOLOLOL" or "hahaha").

A threshold of 10 tweets per ZCTA was chosen to ensure this average RE would not be dominated by a single outlier tweet. We repeated our analysis using thresholds of 1, 5, and 20 tweets per ZCTA. In all cases the anti-correlation between RE and college graduation rate was found, with a slope between $\beta = -0.10$ and $\beta = -0.15$ at the extremes. This shows that our results are not sensitive to choice in threshold.

**SUMMARY AND CONCLUSIONS**

We have presented a novel use of a traditional readability metric, the Flesch RE formula, combined with a large modern data set, Twitter. The language of Twitter is systematically more sophisticated (lower reading ease score) than that of similar digital short-format mediums, including chat rooms and SMS, and even more so when hashtags are included in tweets.

Our analysis of 17.4 million English language tweets characterizes some of the most fundamental properties of communication with this microblogging service. We have found that 84% of tweets do not contain any hashtags, and that approximately 2% of tweets contain geo-data (an increase from previous studies). The reading ease of geo-tagged tweets is higher (easier to read) than that of tweets in general, which we attribute to increased brevity and abbreviation when typing on mobile devices.

We have also shown a significant anti-correlation between average reading ease and college graduation rate within ZIP Code Tabulation Areas. Using the same data sources, a preliminary look at median household income, as well as racial composition failed to produce any statistically significant correlations. However, we have only studied a small handful of the potentially important demographic statistics available.

Our results point to a real change in the language complexity of tweets correlated with (but not necessarily cause by) education level. This may be due to underlying real-world linguistic differences that trace socio-economics, regional dialects/slang, or user age. Readability may thus be a fruitful method to tailor content on Twitter to a diverse range of users, improving the effectiveness of public discourse. The trend in RE might instead point to differences in the type of content being tweeted, e.g. individual status update versus sharing news stores. A large sample of content-indexed tweets with geo-data is needed to test this hypothesis.